\documentclass[amsfonts,amsmath,prd,twocolumn,nofootinbib,a4paper,eqsecnum,superscriptaddress]{revtex4-2}
\newcommand{\beq}{ \begin{equation}}
\newcommand{\eeq}{\end{equation}}

\newcommand{\bsp}{\begin{split}}

\usepackage{epsfig,bbm,cancel}
\usepackage[normalem]{ulem}
\usepackage[colorlinks=true]{hyperref}
\usepackage{latexsym}
\usepackage[utf8]{inputenc}
\usepackage{amsmath}
\usepackage{amsfonts}
\usepackage{amssymb}
\usepackage{booktabs}
\usepackage{array}
\usepackage{tabularx}
\usepackage{multirow}
\usepackage{longtable}
\usepackage{ragged2e}
\usepackage{braket,booktabs}
\usepackage{braket,amsmath}
\usepackage{xcolor}
\usepackage{graphicx}
\usepackage{orcidlink} 

\begin{document}

\title{Parameter Constraints on $f(R^{1+\epsilon},T_{\mu\nu}T^{\mu\nu})$ Gravity from Exotic Compact Objects}

\author{S. Wilhelmina \orcidlink{0009-0007-2052-0002}}
\email{susan.wilhelmina@ui.ac.id}
\author{B. N. Jayawiguna \orcidlink{0000-0001-5920-8701}}
\email{nugrahabyon312@gmail.com}
\author{A. Sulaksono \orcidlink{0000-0002-1493-5013}\thanks{Corresponding author.}}
\email{anto.sulaksono@sci.ui.ac.id}
\affiliation{Departemen Fisika, FMIPA, Universitas Indonesia, Depok, 16424, Indonesia.}

\begin{abstract}
According to General Relativity (GR), gravity is described as the curvature of spacetime caused by mass and energy. Although GR has been highly successful in describing gravitational phenomena, exotic compact objects with high mass and strong gravitational fields, as well as low-mass objects, have motivated the study of modified theories of gravity. One of the modified gravity frameworks is $f(R,T^2) = R^{1 + \epsilon} + \alpha \kappa^2 T_{\mu \nu} T^{\mu \nu}$, which provides an extension of GR by introducing modifications to both the geometric and matter sectors. In this study, we investigate the mass-radius structure, physical viability, and stability of quark stars in $f(R,T^2)$ gravity using the MIT Bag Model equation of state. We compare results from the exact $f(R,T^2)$ model, its first-order Taylor expansion, and GR. The values of the gravity parameter $\epsilon$ and $\alpha$ are constrained by evaluating the ability of the model to reproduce the observed mass-radius configurations of compact objects, including HESS J1731-347 with a mass of $M = 0.77_{-0.17}^{+0.20} \, M_{\odot}$ and a radius of $R = 10.4_{-0.78}^{+0.86}$ km, GW190814 with a mass of $M = 2.6 \, M_{\odot}$, PSR J1614-2230, with a mass of $M = 1.928 \pm 0.017 \, M_{\odot}$ and a radius of $R > 10$ km, and PSR J0437-4715, with a mass of $M = 1.418 \pm 0.037 \, M_{\odot}$ and a radius of $R = 11.36_{-0.63}^{+0.95}$ km. We assess the physical viability and stability of the stellar configurations using the energy conditions (strong, weak, null, and dominant) and the speed of sound. Overall, the $f(R,T^2)$ gravity model can accommodate quark star configurations and reproduce the observed properties of massive compact objects, particularly those associated with GW190814 and the lightest compact object observed, HESS J1731-347, respectively.

\vspace{0.2 cm}

\end{abstract}

\maketitle
\thispagestyle{empty}
\setcounter{page}{1}

\section{Introduction}
LIGO/Virgo observation of the GW 190814 event from the black hole merger with a secondary compact object with $ M = (2.50-2.67) \, M_\odot$ is still physically interesting, related to the question of whether exotic compact objects exist between the maximum mass of neutron stars and the minimum mass of black holes. Please see Refs.~\cite{LIGOScientific:2020zkf,Zhang:2020zsc} and the references therein. The nature of this secondary compact object is unknown, as the signal showed no measurable tidal effects and no electromagnetic counterpart was identified. It is therefore unclear whether the secondary compact object is a black hole, a neutron star, or another exotic object. Several proposals explain this object's nature. Please see Refs.~\cite{LIGOScientific:2020zkf,Zhang:2020zsc,Neves:2024ggn,Prasetyo:2021kfx} and the references therein. On the other hand, a recent measurement of HESS J1731 reports $ M=0.77^ {+0.20} _ {-0.17} M_\odot$ and $ R=10.4^ {+0.86} _ {- 0.78} $ km~\cite{Doroshenko:2022nwp}, which challenges our understanding of compact object matter at low densities. Supernova simulations predict $M=1.17 M_\odot$ as the mass of the lightest possible neutron stars. In contrast, theoretical calculations in general relativity (GR) predict neutron-star masses of $M=(0.88-1.28)M_\odot$. See Ref.~\cite{Sagun:2023rzp} for the discussion of the uncertainty for HESS J1731-347. They found, from joint analysis of thermal evolution and mass-radius constraints, that HESS J1731-347 could be a neutron star with a soft equation of state, a strange quark star, a hybrid star with an early deconfinement phase transition, or a dark matter-admixed neutron star.   Therefore, it is physically interesting to investigate a theory that can explain both exotic objects simultaneously.

Beyond black holes and neutron stars, researchers have proposed several hypothetical compact objects. Their properties remain largely unexplored, and their existence in nature has not been confirmed. These are called exotic compact objects. They represent equilibrium configurations in the strong-gravity regime and involve novel physical phenomena closely connected to fundamental questions in theoretical physics. Quark stars fall into this category~\cite{Bezares:2024btu}. Quark stars represent a theoretical "third state" of stellar death—denser than neutron stars but not quite black holes—providing a unique laboratory to test fundamental physics at extreme densities. They are hypothesized to be composed of "strange quark matter," a deconfined soup of quarks that could be the universe's most stable form of matter \cite{Kumari:2021tik}. Please see the recent review of quark stars and the corresponding equation of state (EoS) in Refs.~\cite{Zhang:2024xod,Danarianto:2024vih} for details. The MIT Bag model is the simplest equation-of-state (EoS) model for quark matter. This model provides a simplified framework for describing quark and gluon confinement within protons and neutrons. It posits that quarks and gluons determine the system's lowest-energy state, plus a vacuum energy density inside the region that arises from the vacuum outside this region; the vacuum energy, known as the bag constant, represents the energy within the region that confines quarks and gluons. The bag constant also generates a confinement force, balancing the internal pressure of the quarks with the external vacuum pressure. This framework allows the construction of an equation of state that includes a vacuum-energy contribution. Although the MIT Bag model is a phenomenological approach and does not capture the detailed interactions of quantum chromodynamics (QCD), it provides qualitative insights into the behavior of matter under extreme density conditions~\cite{Neves:2024ggn}.

General Relativity (GR) is widely considered the most successful theory of gravity because of its conceptual simplicity and its ability to account for a broad range of experimental and observational tests, from ground-based measurements to cosmic-scale phenomena. However, GR does not provide a universally applicable framework for studying gravity. Theoretical arguments and empirical evidence suggest that modifications to GR are required in the strong-gravity (ultraviolet, UV) and large-distance (infrared, IR) regimes. Further details are explained in Ref.~\cite{Shankaranarayanan:2022wbx,Berti:2015itd,Blazquez-Salcedo:2022dxh}. One of the simplest ways to modify GR is by incorporating higher-order curvature invariants into the action. $f(R)$ theories belong to this class of modified gravity. The reviews of $f(R)$ gravity theory can be consulted in Refs.~\cite{Sotiriou:2008rp,DeFelice:2010aj,Nojiri:2017ncd}. Another possibility is a modification in the matter sector.  For example, the energy-momentum squared gravity (EMSG) \cite{Cipriano2024EMSG,Alam:2023grx}.
EMSG is a modification of GR in the matter sector that allows a self-contraction of the energy-momentum tensor $T^{\mu\nu}$, and thereby one can incorporate in the generic functional action an additional term of the form
$f(T_{\mu\nu}T^{\mu\nu}) = \alpha T_{\mu\nu}T^{\mu\nu}$, where
$\alpha$ is a parametric constant. Quark stars have been used as test beds for modified gravities. Therefore, many studies have examined modified gravity using quark-matter EoS. For example, Bumblebee gravity (see Ref.~\cite{Neves:2024ggn} and references therein), Einstein–Gauss–Bonnet gravity (see Ref.~\cite{Gammon:2024gij,Pretel:2021gqq} and references therein), and many other modified gravity theories. For example, see Refs.~\cite{Danarianto:2024vih,Tangphati:2026jid,Tangphati:2024war,Haghani:2023nrm} and the references therein. Studies of quark stars in $f(R)$ theory appear in Refs.~\cite{Astashenok:2014dja,Katayama:2025mok,Pretel2022Anisotropic,HarkoLobo2019}, while studies of quark stars in EMSG theory include Refs.~\cite{Banerjee:2025ewc,Nari:2018aqs} and the references therein. A comprehensive study of neutron stars in $f(R)$ theory appears in Ref.~\cite{Nobleson:2023wca}, while the EMSG case appears in Ref.~\cite{Alam:2023grx}.
See Ref.~\cite{Olmo:2019flu} for a review of compact objects in modified gravity theories.

In the present work, we study whether quark stars within $f(R^{1+\epsilon}, T_{\mu\nu}T^{\mu\nu})$ based on the MIT Bag Model equation of state can explain two extreme or exotic compact objects, such as the secondary object of the GW190814 event and  HESS J1731, simultaneously. We also investigate the mass-radius structure, physical viability, and stability of quark stars in $f(R,T^2)$ gravity. We compare results from the exact model, its first-order Taylor expansion, and GR. We also use the universal prescription for the matter sector of the Lagrangian density, as introduced in Ref. \cite{Jayawiguna:2026das}. The reason for choosing the $f(R) = R^ {1+\epsilon} $ form in the geometrical part of the gravity theory is that, to explain the flat galactic rotation curves and increasing maximum mass of compact objects, only small deviations from GR are needed, and  $R^{1+\epsilon}$  is a quite natural modification to GR in the gravity context in general. The details are in Refs.~\cite{Pretel2022Anisotropic,Jayawiguna:2026ves}. The realistic speed of sound of compact-object matter is known to be radius-dependent. Therefore, to improve the constant speed of sound of quark matter within the MIT Bag model, we included the   $T_{\mu\nu}T^{\mu\nu}$ term in the Lagrangian density of the model considered in this work.
Throughout this work, we use geometrized units, $G=c=1$. Therefore, we express all physical quantities in terms of length.

This paper is structured as follows. In Section II, we derive the field equations for curvature-matter terms, $f(R^{1+\epsilon}, T_{\mu\nu}T^{\mu\nu})$, and examine the second-derivative issue. This section also discusses the numerical procedure for calculating the mass-radius profiles with the MIT bag model. In Section III, we present several validity parameters of the model, such as compactness, speed of sound, and the formalism of the energy conditions in generalized curvature matter. In Section IV, we show all numerical results and discussions. Lastly, we present the conclusion in Section V.

\section{Field Equations}
In this section, we introduce the theoretical framework describing the gravitational field and matter content of our model. We begin by briefly reviewing the theory's gravitational and matter sectors and specifying the action. In particular, we consider $f(R,T^2)$ gravity, in which the gravitational action extends the conventional $f(R)$ form by adding a term constructed from the energy-momentum tensor. This additional coupling lets matter directly modify gravitational dynamics and provides a framework for investigating deviations from GR. The action governing our model can be written as
\begin{equation}
S = \frac{1}{2 \kappa^2} \int \sqrt{-g} \, f(R,T^2) \, \, d^4x + \int \, \mathcal {L}_m \, \sqrt{-g} \, \, d^4x,
\end{equation}
where $\kappa^2=8\pi$, $g$ is the determinant of the metric tensor, and $\mathcal{L}_{m}$ is the Lagrangian in the matter sector. From now on, we define the model from the matter sector to be squared gravity

\begin{eqnarray}
f(R,T^2) = f(R) + \alpha \kappa^2 T_{\mu\nu}T^{\mu\nu},
\end{eqnarray}
while we keep the pure gravity part for later. Using the least action principle in terms of the metric tensor, we could get the modified field equations written as

\begin{eqnarray}
\label{EoM1}
&& f_R R_{\mu \nu} - \frac{1}{2} g_{\mu \nu} f + (g_{\mu \nu} \Box - \nabla_{\mu} \nabla_{\nu}) f_R = \kappa^2 T_{\mu \nu} \nonumber \\ && + \frac{1}{2} \alpha \kappa^2 ( g_{\mu \nu }T_{\sigma \epsilon} T^{\sigma \epsilon} - 2 \Theta^{T^2}_{\mu \nu}).
\end{eqnarray}

where $f_{R} \equiv \partial f/\partial R,$ $g_{\mu\nu}$ is the metric tensor, $f\equiv f(R)$, $\Box\equiv \nabla_{\mu}\nabla^{\mu}$, $T_{\mu\nu}$ is the energy momentum tensor, $f_{T^2} \equiv \partial f/\partial T^2$, and
\begin{eqnarray}
\label{auxuliary}
\Theta^{T^2}_{\mu \nu} &\equiv& \frac{\delta T^2}{\delta g^{\mu \nu}}, \nonumber \\ &=& 2 T_{\alpha \nu} T^{\alpha}_{\mu} + 2 T^{\alpha \beta} \left(\frac{\delta T_{\alpha \beta}}{\delta g^{\mu \nu}}\right)\nonumber \\
&=& 2 T_{\alpha \nu} T^{\alpha}_{\mu} + (-2 T_{\mu \nu} + T \, g_{\mu \nu}) \mathcal{L}_m - T \, T_{\mu \nu} \nonumber \\ &&- 4 T^{\alpha \beta} \left(\frac{\delta^2 \mathcal{L}_m}{\delta g^{\mu \nu} \delta g^{\alpha \beta}}\right).
\end{eqnarray}
The quantity $T\equiv g_{\mu\nu}T^{\mu\nu}$ is the trace of the energy-momentum tensor. From the field expression \eqref{EoM1}, the $f(R)$ gravity case can be obtained by switching off the square energy momentum parameter, $\alpha = 0$.

In order to evaluate further, we employ the standard isotropic perfect fluid for modelling the internal structure of the star
\begin{eqnarray}
T_{\mu \nu} = (\rho + p)u_\mu u_\nu + p g_{\mu \nu},
\end{eqnarray}
where $\rho$, $p$, and $u_{\mu}$ are density, pressure, and 4-velocity, respectively. The 4-velocity is normalized such that $u_{\mu}u^{\mu}=-1$. The additional term from coupling curvature and matter comes from $\mathcal{L}_{m}$ and its second derivative. In particular, matter density and radial pressure often serve as the matter Lagrangian density because both are thermodynamically consistent and yield the same Einstein field equations (GR). New gravity models with explicit links between matter and geometry instead involve second-order derivatives of $\mathcal{L}_{m}$ relative to the metric tensor. So, picking either $p$ or $-\rho$ for $\mathcal{L}_{m}$ gives different effective field equations. This confusion appears because one usually treats the four-velocity ($u_{\mu}$) and the metric tensor ($g_{\mu\nu}$) as independent. In \cite{Jayawiguna:2026das}, the authors offer an alternative framework by relaxing that assumption, thereby making the modified gravity theory independent of the choice of $\mathcal{L}_{m}$. We will use this method for the rest of this paper. However, the second derivative \cite{Jayawiguna:2026das} can be written in compact as 
\begin{eqnarray}
\frac{\delta^2 p}{ \delta g^{\mu \nu} \delta g^{\alpha \beta} }&=& - \frac{(\rho+p)}{4}u_{\mu}u_{\nu}\left( g_{\alpha\beta} +2 u_{\alpha}u_{\beta} \right),\\ \frac{\delta^2 \rho}{ \delta g^{\mu \nu} \delta g^{\alpha \beta} }&=&- \frac{\delta^2 p }{\delta g^{\mu\nu}\delta g^{\alpha\beta}} + \frac{(\rho+p)}{4}g_{\alpha\beta}g_{\mu\nu}\nonumber \\&& -\frac{(\rho+p)}{2} g_{\mu\alpha} g_{\nu\beta}.
\end{eqnarray}
Thus, $\Theta^{T^2}_{\mu\nu} (\mathcal{L}_m=p) = \Theta^{T^2}_{\mu\nu}(\mathcal{L}_{m}=-\rho).$ These choices make the $\Theta^{T^2}_{\mu\nu}=0,$ and hence, the field equations reduce to
\begin{eqnarray}
&& f_R R_{\mu \nu} - \frac{1}{2} g_{\mu \nu} f + (g_{\mu \nu} \Box - \nabla_{\mu} \nabla_{\nu}) f_R = \nonumber \\&&  \kappa^2 \rho \Big[\left(1 + \frac{p}{\rho}\right) u_{\mu} u_{\nu} + \frac{p}{\rho} g_{\mu \nu}\Big] + \frac{1}{2} \alpha \kappa^2 \rho^2 \left(1 + \frac{3p^2}{\rho^2}\right) g_{\mu \nu},
\end{eqnarray}
these terms can be denoted as effective energy-momentum tensor, such that the equation becomes
\begin{equation}
\label{eomsimplified}
f_R R_{\mu \nu} - \frac{1}{2} g_{\mu \nu} f + (g_{\mu \nu} \Box - \nabla_{\mu} \nabla_{\nu}) f_R = \kappa^2 T^{\mathrm{eff}}_{\mu \nu},
\end{equation}
with an effective energy-momentum tensor given by $T^{\mathrm{eff}}_{\mu \nu} = (\rho^{{\mathrm{eff}}} + p^{\mathrm{eff}}) u_{\mu} u_{\nu} + p^{\mathrm{eff}} g_{\mu \nu}$, the resulting effective energy density and effective pressure are given by the following equations
\begin{align}
\rho^{\mathrm{eff}} &= \rho - \frac{1}{2} \alpha \rho^2 \left(1 + \frac{3p^2}{\rho^2  }\right), \\
p^{\mathrm{eff}} &= p + \frac{1}{2} \alpha \rho^2 \left(1 + \frac{3p^2}{\rho^2}\right).
\end{align}
Eq. \eqref{eomsimplified} provides a complete, simplified form of the curvature-matter coupling for energy-momentum squared gravity. We discuss each component in detail using the static, spherically symmetric ansatz metric below.

\begin{equation}
\label{ansatz}
ds^2 = -e^{2 \nu (r)} dt^2 + e^{2 \lambda (r)} dr^2 + r^2 d\theta^2 + r^2 \sin^2 \theta d\phi^2,
\end{equation}
and the field equations are given in terms of the effective matter part
\begin{widetext}
	\begin{eqnarray}
	\label{00_2}
	-\frac{f_R}{r^2} + \frac{f_R}{r^2} \frac{d}{dr} (re^{-2 \lambda}) + \frac{1}{2}(R f_R -f) + e^{-2 \lambda} \left[\left(\frac{2}{r} - \lambda'\right)f'_R + f''_R\right] = - \kappa^2 \rho^{\text{eff}} ,
	\end{eqnarray}
	\begin{eqnarray}
	\label{11_2}
	-\frac{f_R}{r^2} + \frac{f_R}{e^{2 \lambda}} \left(\frac{2 \nu'}{r}+ \frac{1}{r^2}\right) + \frac{1}{2} (R f_R - f) + e^{- 2 \lambda} \left(\frac{2}{r} + \nu'\right) f'_R = \kappa^2 p^{\text{eff}} ,
	\end{eqnarray}
	\begin{equation}
	\label{22_33_2}
	\frac{f_R}{r^2} \left[1 + e^{-2 \lambda} (r \lambda' - r \nu' - 1)\right] - \frac{1}{2} f + e^{- 2 \lambda} \left[\left(\frac{1}{r} + \nu' - \lambda'\right) f'_R + f''_R\right] = \kappa^2 p^{\text{eff}}.
	\end{equation}
\end{widetext}
Setting $f(R)=R$ and $\alpha=0$ make the equation above into standard GR with perfect fluid.

Since we have not specified $f(R)$, the Ricci scalar, $R$, and metric tensor are now dynamical fields, and the relation for solving the Ricci scalar can be obtained by taking the trace of Eq. \eqref{eomsimplified}
\begin{equation}
\label{trace2}
3 \Box f_R + R f_R - 2f = \kappa^2 \, T^{\text{eff}},
\end{equation}
and therefore the equation becomes
\begin{eqnarray}
\label{dynamicfinal}
&& \frac{3}{e^{2 \lambda}} \Big[\left(\frac{2}{r} + \nu' - \lambda'\right) f'_R + f''_R\Big] = \kappa^2 \rho \left(-1 + \frac{3p}{\rho}\right) \nonumber \\ && + \frac{1}{2} \alpha \kappa^2 \Big[4 \rho^2 \left(1 + \frac{3p^2}{\rho^2}\right)\Big] + 2f - R f_R .
\end{eqnarray}

The stellar mass function, along with all differential equations governing the star's interior structure, can be derived from the non-zero components of the $f(R,T^2)$ gravity field equations \eqref{00_2}-\eqref{22_33_2} and the dynamical equation \eqref{dynamicfinal}. We obtain the mass function for the stellar fluid from Eq. \eqref{00_2} using $e^{-2\lambda}=1-2m(r)/r$ in the field equation.
\begin{eqnarray}
m(r) &=& 4 \pi \int^{r}_{0} r'^2 \rho^{\mathrm{eff}} (r') dr' \nonumber \\ && + \frac{1}{2} \int^{r}_{0} \Big\{\frac{(1- f_R)}{r'^2} \frac{d}{dr} [r'(1 - e^{-2 \lambda})] \nonumber \\ && + \frac{1}{e^{2 \lambda}} \Big[\left(\frac{2}{r'} - \lambda'\right) R' f_{RR} + R'' f_{RR} + R'^2 f_{RRR}\Big]  \nonumber \\ &&+ \frac{1}{2} (R f_R - f)\Big\} r'^2 dr',
\end{eqnarray}
while in GR case, the second integral vanishes.

In closed form, we construct the system of equations from Eq. \eqref{00_2} to obtain the equation for $\lambda'$, and from Eq. \eqref{11_2} to obtain the equation for $\nu'$. From the conservation equation, $\nabla_{\nu}T^{\nu}_{\mu ~\textrm{eff}}=0,$ we can get the effective TOV equation. This modified TOV equation serves as the foundation for obtaining solutions of compact stars with isotropic pressure under static and relativistic conditions. Lastly, we obtain the second-order derivative of the Ricci scalar curvature $R$ from the dynamical equation in Eq. \eqref{dynamicfinal}. The final equations are as follows
\begin{widetext}
	\begin{align}
	\nu' &= \frac{1}{2r (2 f_R + r R'f_{RR})} [r^2 e^{2 \lambda} (16 \pi p^{\mathrm{eff}} + f - R f_R) + 2 f_R (e^{2 \lambda} - 1) - 4r R'f_{RR}], \\
	\lambda' &= \frac{1}{2 r \left(2 f_R + r R' f_{RR}\right)} \Bigg\{2 f_R (1 - e^{2 \lambda}) + 2 r^2 e^{2 \lambda} 8 \pi \rho^{\mathrm{eff}} \nonumber \\
	& \qquad + \frac{r^2 e^{2 \lambda}}{3} \Big[16 \pi \rho \left(-1 + \frac{3p}{\rho}\right) + \frac{1}{2} 2 \alpha \kappa^2 \left(4 \rho^2 \left(1 + \frac{3p^2}{\rho^2}\right)\right) + R f_R + f\Big] \nonumber \\
	& \qquad + \frac{r R' f_{RR}}{f_R} \Big[2 f_R (1 - e^{2 \lambda}) + r^2 e^{2 \lambda} 8 \pi \rho^{\mathrm{eff}} + \frac{r^2 e^{2 \lambda}}{3} \Big[8 \pi \rho \left(-1 + \frac{3p}{\rho}\right) + \frac{1}{2}  \alpha \kappa^2 \left(4 \rho^2 \left(1 + \frac{3p^2}{\rho^2}\right)\right) \nonumber \\
	&\qquad - 24 \pi p^{\mathrm{eff}} + 2 R f_R - f \Big] + 2r R' f_{RR} \Big] \Bigg\}, \\
	\frac{d^2R}{dr^2} &= \frac{1}{3 f_{RR}} \Big\{e^{2 \lambda} \Big[\kappa^2 \rho \left(-1 + \frac{3p}{\rho}\right) + \frac{1}{2} \alpha \kappa^2 \Big[4 \rho^2 \left(1 + \frac{3p^2}{\rho^2}\right)\Big] + 2f - R f_R\Big] - 3 R'^2 f_{RRR}\Big\} \nonumber \\ 
	& \quad + \left(\lambda' - \nu' - \frac{2}{r}\right) R', \\
	\frac{dp}{dr} &= - \frac{1}{1 + 3 \alpha p} \Big[(\rho+ p) \nu' + \alpha \rho \frac{d \rho}{dr}\Big].
	\end{align}
\end{widetext}

The TOV equation captures the role of gravity, mechanical equilibrium, and hydrodynamics in determining the structure of relativistic stars. The equations for $m', \, \nu', \,\text{and} \,\, p'$ represent a star that is spherically symmetric, static, and relativistic \cite{Glendenning1997CompactStars, Ryder2009GR,HarkoLobo2019}. With the specified $f(R)$ form, we can simultaneously solve these coupled equations with the specified EoS, along with the imposed boundary conditions in the near center as
\begin{eqnarray}
&& p(0)=p_{c},~~~~\nu(0)=\nu_c,~~~~\lambda(0)=0,\nonumber \\ &&R(0)=R_{c},~~~~\textrm{and}~~~~ R'(0)=0,
\end{eqnarray}
We have five variables to determine; solving the system of differential equations requires boundary conditions, including setting the mass at the stellar center to zero. The integration is carried out starting from a small radius $r = r(0) $, where the condition can be obtained, and the surface of the star,~$p(r_s)=0$, and the total gravitational mass is evaluated at that radius, $r_{s}$. The numerical integration begins from a very small radius to represent the conditions at the stellar center. In the implementation, we use an initial value of $r_0 = 10^{-5}$ to avoid the numerical singularity at $r = 0$ caused by division by zero. Since the integration starts at a small radius, it proceeds outward from the stellar center until the pressure satisfies $p(r)/p_c = 10^{-9}$, which marks the stellar surface boundary condition. The gravitational mass is determined using the Schwarzschild metric
\begin{equation}
M(r) = \frac{r}{2} (1 - e^{-2 \lambda}).
\end{equation}
This expression is evaluated at the stellar surface. We vary the central pressure iteratively to obtain different stellar configurations. Each value of the central pressure generates a single mass–radius (M-R) pair. By repeating the process, the M-R curve for the quark stars is obtained; we will discuss it in the results and discussion.

However, the star profile outside the star is simpler since we no longer need pressure and density ($\rho=p=0$). The equations reduce to

\begin{widetext}
	\begin{eqnarray}
	\frac{d\nu}{dr} &=&    \frac{r^2e^{2\lambda}(f-Rf_{R})  +2f_{R} (e^{2\lambda}-1) -4r R' f_{RR}}{2r(2f_{R}+rR'f_{RR})}  ,  \\ \frac{d\lambda}{dr} &=&   \frac{1}{2r(2f_{R}+rR' f_{RR})} \bigg \lbrace 2f_{R} (1-e^{2\lambda}) +\frac{r^2 e^{2\lambda}}{3} [Rf_{R}+f]    \nonumber \\ && +\frac{rR' f_{RR}}{f_{R}} \left[ 2f_{R} (1-e^{2\lambda}) +\frac{r^2 e^{2\lambda}}{3} (2Rf_{R}-f)+2rR'f_{RR}    \right]  \bigg \rbrace       ,   \\ \frac{d^2 R}{dr^2} &=&  \frac{1}{3f_{RR}} \bigg\lbrace e^{2\lambda} [2f-Rf_{R}] -3R'^2 f_{RRR} \bigg\rbrace +\left( \lambda'-\nu'-\frac{2}{r} \right) R',
	\end{eqnarray}
\end{widetext}
Moreover, to get the complete solution, we have to make sure that the solution covers both interior and exterior by matching the solution with junction condition at the star surface
\begin{eqnarray}
\label{conditions}
[\nu]=0,~~~[\lambda]=0,~~~[R]=0,~~~[R']=0,
\end{eqnarray}
Here, $[x]$ represents the difference between the value of $x$ evaluated inside and outside at $\Sigma_{0}$, specifically $[x]=x^{-}|_{\Sigma_{0}}-x^{+}|_{\Sigma_{0}}$. In this convention, $+$ and $-$ denote quantities outside and inside the star, respectively, with $\Sigma_0=r_s$. The influence of $f(R)$ is not confined to the surface; consequently, it persists. The exact Schwarzschild condition is recovered by choosing $f(R)=R$. In the generalized context, the Schwarzschild condition arises from the requirement of asymptotic flatness.
\begin{eqnarray}
\lim_{r\rightarrow \infty} R(r)=0,~~~~  \lim_{r\rightarrow \infty} m(r)=\textrm{constant},~~~~\lim_{r\rightarrow \infty} \nu(r)=0,
\end{eqnarray}
means that $ R_c$ and $\nu_c$ must be chosen so that the requirement above is satisfied at infinity.

In fact, obtaining such a solution requires specifying the particular model of $f(R)$ gravity. An interesting class of models is the power-law models \cite{Pretel2022Anisotropic,Capozziello2016}. The exact form and its derivative can be shown as
\begin{align}
f &= R^{1+ \epsilon}, \\
f_R &= (1+ \epsilon) R^{\epsilon}, \\
f_{RR} &= \epsilon (1+ \epsilon) R^{\epsilon - 1}; \\
f_{RRR} &= \epsilon (\epsilon - 1) (1+ \epsilon) R^{\epsilon - 2},
\end{align}
Here, $\epsilon$ is a free parameter measuring the deviation of the gravitational theory from GR. We choose $|\epsilon| \ll 1$ because we are interested in a modified gravity theory that remains close to GR. In the limit $\epsilon \rightarrow 0$, the equation will reduce to the Einstein-Hilbert form, namely $f(R)=R$. Furthermore, in addition to proposing the alternative model $f(R^{1+\epsilon}, T^2)$, this study compares numerical results obtained using the exact form of $f(R)$ with those obtained using the first-order Taylor expansion of $f(R)$. The first-order Taylor-expanded form of $f(R)$ and its derivatives are given as follows.
\begin{align}
f &\approx R + \epsilon R \ln{R}, \label{ftaylor}\\
f_R &= 1 + \epsilon + \epsilon \ln{R}, \label{fRtaylor}\\
f_{RR} &= \frac{\epsilon}{R}, \label{fRRtaylor}\\
f_{RRR} &= - \frac{\epsilon}{R^2} \label{fRRRtaylor}. 
\end{align}
The choice of $\epsilon$ close to zero is also motivated by weak-field tests of gravity. In weak gravitational fields, such as those probed by Solar System observations, constraints tend to require $\epsilon$ to be close to zero. In this work, we consider $\epsilon < 0$ because it yields interesting physical results \cite{Pretel2022Anisotropic,Capozziello2016}. However, compact objects have much stronger spacetime curvature, making them useful for investigating whether modified gravity effects can arise when $\epsilon$ slightly deviates from zero. The Taylor-expanded form given in Eqs.(\ref{ftaylor}--\ref{fRRRtaylor}) has been widely studied, from the Solar System scale to cosmological scales \cite{Capozziello2016,ASTASHENOK2020135910}. These studies have shown interesting results, including applications to analyzing the cosmological background of gravitational waves and to black hole solutions using Noether symmetry, where this Taylor-expanded form is employed \cite{Capozziello2016}. Therefore, this study also investigates the numerical deviation between the results obtained from the exact form and those obtained from the Taylor-expanded form for $|\epsilon| \ll 1$. In particular, we aim to determine whether small deviations from GR can significantly change the structure and physical properties of compact stars. This is because, in the first-order Taylor expansion, a smaller $\epsilon$ yields a smaller deviation from the exact form.

As with the construction of isotropic compact stars, closing the system requires specifying an EoS. We employ the MIT bag model EoS for the dense matter involved in the quark star. We express the equations in terms of energy density as a function of pressure.
\begin{equation}
\rho(r) = 3p(r) + 4B,
\end{equation}
where $B$ is the bag constant; the bag constant used for this work is $60$ \text{MeV/fm$^3$}. In geometrized units, the bag constant is $7.943 \times 10^{-5}~ \text{km}^{-2}$.

The $f(R,T^2)$ theory extends $f(R)$ gravity by introducing the energy-momentum term $T^2$, which modifies the right-hand side of the field equations. In regions of strong spacetime curvature, this additional contribution can deviate from both GR and $f(R)$ gravity, affecting the internal structure and mass-radius relation of compact objects. It may also influence their stability, the final state following gravitational collapse \cite{Cipriano2024EMSG}, and constrain the allowed parameter range, as discussed in the following section.

\section{Validity Parameter of the model}

\subsection{The compactness and speed of sound}

The important quantity related to stellar stability is the compactness. Compactness is a dimensionless quantity that characterizes how compact an object is. According to the fundamental theorem proposed by Buchdahl in 1959, the compactness of a static, spherically symmetric, and isotropic object within the GR framework is bounded by 
\begin{equation}
C \equiv \frac{GM}{Rc^2} \leq \frac{4}{9} \approx 0.444,
\end{equation}
where $M$ and $R$ represent the mass and radius of the compact object, respectively \cite{alho2022compactness}. The value $0.444$ is the Buchdahl limit for isotropic, static, and spherically symmetric compact objects \cite{alho2022compactness}.

The speed of sound is also important for determining stability, as it provides information on the stiffness of the equation of state \cite{Lopes2025VectorMITBag}. The speed of sound is defined as
\begin{equation}
c_s^2 = \frac{dp}{d \rho},
\end{equation}
This indicates the pressure response to changes in energy density. To remain within the stable range, the speed of sound must satisfy the following physical condition.
\begin{equation}
0 \leq c_s^2 \leq 1,
\end{equation}
the causality limit is $c_s^2 \leq 1$, while $c_s^2 \geq 0$ ensures thermodynamic stability \cite{BedaqueSteiner2015}. A system is considered stable in terms of the speed of sound if it satisfies the causality condition and thermodynamic stability \cite{BedaqueSteiner2015}. The causality limit is $c_s^2 = 1$, and the stiffest possible equation of state occurs when the speed of sound equals the speed of light \cite{Glendenning1997CompactStars}. 
This research uses the effective speed of sound given by
\begin{equation}
c_s^{2 \, \mathrm{eff}} = \frac{dp^{\mathrm{eff}}}{d\rho^{\mathrm{eff}}},
\end{equation}
because it employs modified gravity with two free parameters, namely $\epsilon$ and $\alpha$, the star is considered stable if $c_s^{2 \, \mathrm{eff}} \in [0,1] $. Note that in GR, the speed of sound is strongly related with the compactness. However, in this model, the parameters $\epsilon$ and $\alpha$ can control the sound speed, keeping it within range, and the compactness is relatively large as well.

\subsection{Energy conditions} \label{subsect: ecs}

In spacetime theory, Energy Conditions (ECs) restrict the energy-momentum tensor to ensure that energy remains positive. Physically, the energy density in the energy-momentum tensor must be non-negative, a necessary condition for systemic stability \cite{Wald1984, HawkingEllis1973,Kontou2020EnergyConditions,Curiel2014}. ECs serve as parameters that keep the physical model realistic; beyond ensuring the energy-momentum tensor remains positive, they guarantee that gravity remains attractive and maintain the consistency of geodesics within that spacetime \cite{Arora2024EnergyConditions}.

The foundation of ECs lies in the Raychaudhuri equation under attractive gravity. The Raychaudhuri equation is defined as follows.
\begin{equation}
\frac{d\theta}{d \tau}=-\frac{1}{3} \theta^2 -\sigma_{\mu \nu} \sigma^{\mu \nu} + \omega_{\mu \nu}  \omega^{\mu \nu}-R_{\mu \nu} u^{\mu} u^{\nu}  ,
\end{equation}
$u^{\mu}$ is the tangent vector field to a timelike geodesic congruence within a spacetime manifold, $\theta$ denotes the expansion parameter, $\sigma^{\mu\nu}$ represents the shear tensor, and $\omega_{\mu \nu}$ represents the rotation parameter. Since the Einstein field equations relate the Ricci tensor $R_{\mu\nu}$ to the energy-momentum tensor $T_{\mu\nu}$, combining Einstein's and Raychaudhuri's equations yields the corresponding expressions for the Energy Conditions (ECs) \cite{Santos2007EnergyConditions}. For the equation governing the expansion of lightlike (null) geodesics, the vector field is represented by $k^{\mu}$. The equation takes a similar form to the standard Raychaudhuri equation, except that the vector field $u^{\mu}$ is replaced by $k^{\mu}$ and with
a factor $1/2$ rather than $1/3$ \cite{Santos2007EnergyConditions}. Since the shear is a spatial tensor, $\sigma_{\mu \nu} \sigma^{\mu \nu} \geq 0$.  Assuming a non-rotating geodesic congruence, the rotational term satisfies  $\omega_{\mu \nu} \omega^{\mu \nu} = 0$, while $\theta^2$ remains invariably positive. The condition for attractive gravity is satisfied when $d\theta/d\tau < 0$ \cite{Arora2024EnergyConditions}. Attractive gravity implies geodesic focusing. Under these assumptions, the following expression is obtained 
\begin{equation}
R_{\mu \nu} u^{\mu} u^{\nu} \geq 0,
\end{equation}
which represents the timelike convergence condition, one also finds
\begin{equation}
R_{\mu \nu} k^{\mu} k^{\nu} \geq 0,
\end{equation}
which represents the null convergence condition. By applying the Einstein Field Equation (EFE), the timelike convergence condition becomes
\begin{equation}
\label{SECGR}
R_{\mu \nu} u^{\mu} u^{\nu} = 8 \pi \left(T_{\mu \nu} - \frac{1}{2}T g_{\mu \nu}\right) u^{\mu} u^{\nu} \geq 0,
\end{equation}
which is known as the Strong Energy Condition (SEC). The SEC is an energy condition related to the attractive nature of gravity \cite{Kontou2020EnergyConditions}. Meanwhile, for the null convergence condition becomes
\begin{equation}
\label{NECGR}
R_{\mu \nu} k^{\mu} k^{\nu} = 8 \pi T_{\mu \nu} k^{\mu} k^{\nu} \geq 0,
\end{equation}
which is known as the Null Energy Condition (NEC). The NEC requires that, for observers moving along lightlike trajectories (null geodesics), the sum of the energy density and pressure must be positive. A non-rotating null geodesic congruence tends to be attractive. This implies that particles traversing these paths remain affected by gravity \cite{Kontou2020EnergyConditions}. For an isotropic perfect fluid in the GR frame, the SEC and NEC equations become \cite{Santos2007EnergyConditions,Arora2024EnergyConditions}. 
\begin{align}
\rho + 3p \geq 0, \\
\rho + p \geq 0.
\end{align}
Another energy condition is the Weak Energy Condition (WEC). In GR, the WEC requires the energy density to remain positive for any observer, in any direction. The physical expression of the WEC in GR is given by \cite{HawkingEllis1973,Kontou2020EnergyConditions}. 
\begin{equation}
\label{WECGR}
T_{\mu \nu} u^{\mu} u^{\nu} \geq 0.
\end{equation}
For the isotropic perfect fluid in GR frame, the WEC equation, respectively, becomes \cite{HawkingEllis1973,Kontou2020EnergyConditions}
\begin{align}
\rho \geq 0, \quad \text{and} \quad \rho + p \geq 0 .
\end{align}
A subsequent energy condition is the Dominant Energy Condition (DEC). The DEC requires that energy must not propagate faster than the speed of light. The DEC condition requires that the energy-momentum flux measured by an observer must be future-directed and cannot propagate faster than the observer’s proper time. This ensures that causality is not violated. The DEC further requires the energy density to dominate, meaning its magnitude must exceed the remaining components of the energy-momentum tensor \cite{Kontou2020EnergyConditions,Curiel2014}. The equation for the DEC is given by
\begin{equation}
T_{\mu \nu} u^{\mu} v^{\nu} \geq 0.
\end{equation}
From the definition, the DEC equation, respectively, becomes \cite{Kontou2020EnergyConditions}
\begin{equation}
\label{DECGR}
T_{00} \geq |T_{\mu \nu}|.
\end{equation}

Since we consider modified gravity, the energy conditions must also be appropriately modified. In the framework of $f(R, T^2)$ gravity, these conditions can be transform into the Einstein frame as in Eq.\eqref{SECGR}. The SEC expression for $f(R, T^2)$ gravity is obtained as follows
\begin{align}
&(\rho^{\mathrm{eff}} + 3 p^{\mathrm{eff}}) + \frac{f}{8 \pi} - \frac{f_R R}{8 \pi} - \frac{2 \nu' \,e^{-2 \lambda} \, R' f_{RR}}{8 \pi} \nonumber \\
&- \frac{e^{-2 \lambda} \Big[\left(\frac{2}{r} + \nu' - \lambda'\right) R' f_{RR} + R'' f_{RR} + R'^2 f_{RRR}\Big]}{8 \pi} \geq 0.
\label{ECY} 
\end{align}
By using the WEC equation as in Eq.\eqref{WECGR}, the WEC expression for $f(R,T^2)$ gravity is obtained as follows
\begin{align}
&\rho^{\mathrm{eff}} - \frac{\frac{1}{2} (f - R f_R)}{8 \pi} - \frac{e^{-2 \lambda} \, \nu' f_{RR} R'}{8 \pi} \nonumber \\
&+ \frac{e^{-2 \lambda} \Big[\left(\frac{2}{r} + \nu' - \lambda'\right) R' f_{RR} + R'' f_{RR} + R'^2 f_{RRR}\Big]}{8 \pi} \geq 0.
\end{align}
By employing the field equations for $f(R, T^2)$ gravity and transforming them into the Einstein frame as in Eq.\eqref{NECGR}, the NEC expression for $f(R, T^2)$ gravity is obtained as follows
\begin{equation}
(\rho^{\mathrm{eff}} + p^{\mathrm{eff}}) + \frac{e^{-2 \lambda} (R'' f_{RR} + R'^2 f_{RRR} - R' f_{RR} (\lambda' + \nu'))}{8 \pi} \geq 0 .
\end{equation}
The DEC expression for $f(R, T^2)$ gravity, derived from the requirement that the energy density must dominate over the other components of the energy-momentum tensor, is obtained as follows
\begin{eqnarray}
&&(\rho^{\mathrm{eff}} - p^{\mathrm{eff}}) - \frac{f}{8 \pi} + \frac{R f_R}{8 \pi} \nonumber \\ && + \frac{e^{-2 \lambda} \Big[\left(\frac{4}{r} - (\lambda' - \nu')\right) R' f_{RR} + R'' f_{RR} + R'^2 f_{RRR}\Big]}{8 \pi} \geq 0.
\label{ECX}
\end{eqnarray}

Energy conditions (ECs) defined in Eqs.~\eqref{ECY}-~\eqref{ECX} are used to assess the physical viability of a star.

\section{Numerical Results}
\subsection{Effect of the Parameters \texorpdfstring{$\epsilon$}{epsilon} and \texorpdfstring{$\alpha$}{alpha} on the Mass-Radius Profile}

This section discusses the mass–radius (M-R) relation of quark stars within the framework of $f(R,T^2) = R^{1+\epsilon} + \alpha \kappa^2 T_{\mu\nu}T^{\mu\nu}$ gravity using the MIT Bag Model. We investigate how the modified gravity parameters, $\epsilon$ and $\alpha$, affect quark-star structure. In addition, we adjust the central pressure $p_c$ to determine the predicted quark-star configurations corresponding to the maximum mass and the stellar radius at that maximum mass. We present the results of these constraints in Table \ref{tab:Table_1}.

\begin{table*}[t]
	\centering
	\resizebox{\textwidth}{!}{%
		\begin{tabular}{|c|c|c|c|c|c|c|}
			\hline
			$\epsilon$ & $\alpha\,[\mathrm{km}^2]$ & Mass [$M_{\odot}$] & $P_c \times 10^{-4}~[\mathrm{km}^{-2}$] & Radius [km] & $C$ & $c_s^{2\,\mathrm{eff}}$ \\
			\hline
			
			$-10^{-6}$ 
			& $0$ 
			& $1.98$
			& $4$ & 10.75 & $0.271$ & $0.33$ \\
			
			\hline
			
			\multirow{4}{*}{$-0.045$}
			& $163$ 
			& $2.62$
			& $4.3$ & 12.49 & $0.309$ & $0.995$ \\
			
			\cline{2-7}
			
			& $163$ 
			& $2.58$
			& $4.3$ & 12.42 & $0.306$ & $0.995$ \\
			
			\cline{2-7}
			
			& $-203$ 
			& $1.58$
			& $3.3$ & 10.89 & $0.213$ & $0.001$ \\
			
			\cline{2-7}
			
			& $-203$ 
			& $1.56$
			& $3.3$ & 10.83 & $0.212$ & $0.001$ \\
			
			\hline
			
			\multirow{4}{*}{$-0.05$}
			& $166$ 
			& $2.67$
			& $4.2$ & 12.64 & $0.311$ & $0.994$ \\
			
			\cline{2-7}
			
			& $166$ 
			& $2.62$
			& $4.2$ & 12.56 & $0.308$ & $0.994$ \\
			
			\cline{2-7}
			
			& $-208$ 
			& $1.59$
			& $3.2$ & 10.96 & $0.213$ & $0.001$ \\
			
			\cline{2-7}
			
			& $-208$ 
			& $1.56$
			& $3.2$ & 10.90 & $0.211$ & $0.001$ \\
			
			\hline
			
			\multirow{4}{*}{$-0.1$}
			& $181$ 
			& $3.09$
			& $3.8$ & 14.00 & $0.325$ & $0.996$ \\
			
			\cline{2-7}
			
			& $181$ 
			& $2.95$
			& $3.8$ & 13.69 & $0.318$ & $0.996$ \\
			
			\cline{2-7}
			
			& $-252$ 
			& $1.68$
			& $2.5$ & 11.70 & $0.212$ & $0.001$ \\
			
			\cline{2-7}
			
			& $-252$ 
			& $1.61$
			& $2.5$ & 11.48 & $0.207$ & $0.001$ \\
			
			\hline
		\end{tabular}%
	}
	\caption{Constrained values of $\epsilon$ and $\alpha$, and the corresponding quark-star properties.}
	\label{tab:Table_1}
\end{table*}

The parameter $\epsilon$ represents the geometric correction to the Ricci scalar, whereas $\alpha$ describes the matter-coupling contribution through $T_{\mu \nu} T^{\mu \nu}$. The parameter $\epsilon$ in the $R^{1+ \epsilon}$ term physically determines the strength of the spacetime curvature response to matter inside the stellar interior. Table \ref{tab:Table_1} provides a global description of the roles of the $\epsilon$ and $\alpha$ parameters. 

\begin{figure*}[t!]
	\centering
	\includegraphics[width=0.49\linewidth]{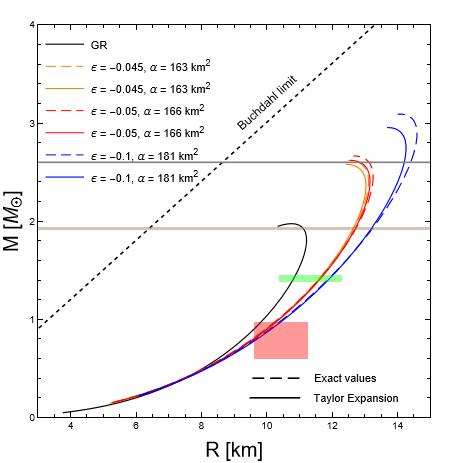}
	\includegraphics[width=0.49\linewidth]{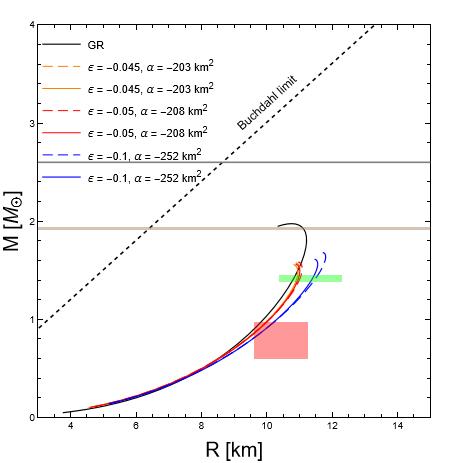}
	\caption{[Left] Mass–radius relations of quark stars for $\epsilon < 0$ and $\alpha > 0$. [Right] Same with left panel, but for $\alpha < 0$. The dashed lines represent the exact model, $f=R^{1 + \epsilon}$, together with its derivatives $f_R, f_{RR}$ and $f_{RRR}$, while the solid lines represent the corresponding first-order Taylor expansion, $f \approx R+\epsilon R\ln R$, and its derivatives. This convention applies to the right-hand panels and all Figures. The red constraints correspond to HESS J1731-347, the green constraints to PSR J0437-4715, the brown constraints to PSR J1614-2230, and the grey constraints to GW190814. The same colour coding is used in Figure \ref{fig:mr-fixed}.}
	\label{fig:perbandingan}
\end{figure*}

Through the geometric correction $R^{1+\epsilon}$, when $\epsilon < 0$, $R$ becomes smaller. Consequently, the geometric contribution changes under the interior conditions of a quark star with high energy density and pressure, since large energy density and pressure produce strong spacetime curvature. More negative $\epsilon$ increases the quark star's maximum mass. In GR, gravity inside compact stars is extremely strong, and the internal pressure eventually fails to sustain the star when the mass becomes too large. In this model, however, the nonlinearity in the $R^{1+\epsilon}$ function modifies hydrostatic equilibrium through additional geometric corrections, making the effective gravitational response to large energy density and pressure weaker than in GR. As a result, internal matter pressure becomes more effective at resisting gravitational collapse, supporting larger stellar masses and allowing quark star configurations with higher maximum masses to form. In other words, when effective gravity weakens and effective pressure increases, the pressure can sustain stars against gravitational collapse, allowing them to maintain equilibrium at higher masses. To investigate the effects of the free parameters $\epsilon$ and $\alpha$ on stellar structure, we also vary each parameter separately and plot the resulting properties.

\begin{figure*}[t!]
	\centering
	\includegraphics[width=0.49\linewidth]{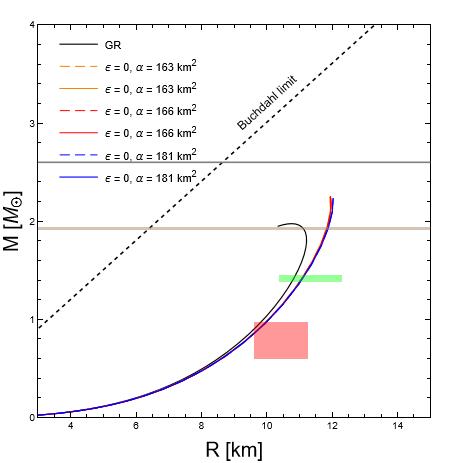}
	\includegraphics[width=0.49\linewidth]{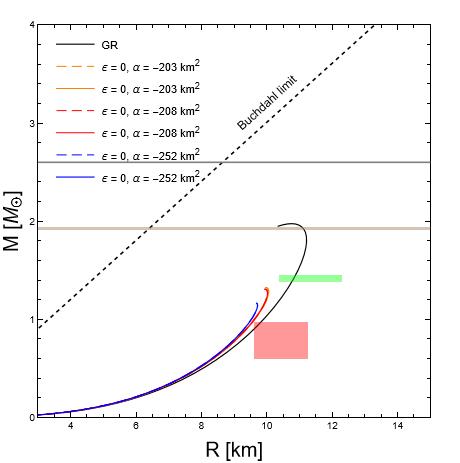}
	\includegraphics[width=0.49\linewidth]{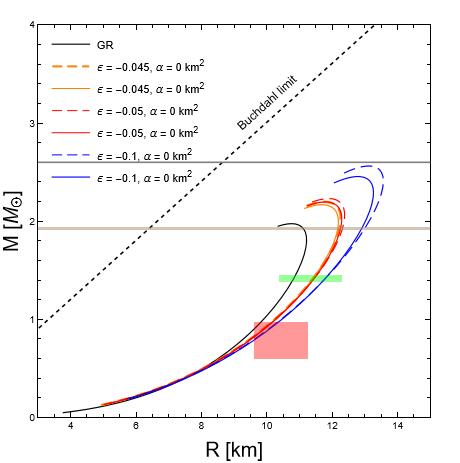}
	\caption{[Top Left] Mass–radius relations of quark stars for $\epsilon = 0$ and $\alpha > 0$. [Top Right] Mass–radius relations of quark stars for $\epsilon = 0$ and $\alpha < 0$. [Bottom Center] Mass–radius relations of quark stars for $\epsilon < 0$ and $\alpha = 0 ~\textrm{km}^2$.}
	\label{fig:mr-fixed}
\end{figure*}

More negative values of $\epsilon$ also lead to a larger stellar radius. As mentioned earlier, when effective gravity weakens, pressure dominates, allowing matter to spread outward and increasing the stellar radius. Because the star becomes larger, it can accommodate more mass before eventually collapsing. In other words, weaker effective gravity leads to a larger radius and mass. This behavior can be seen in Table \ref{tab:Table_1} or Figure \ref{fig:perbandingan}; configurations with negative $\epsilon$ and positive $\alpha$ yield larger stellar masses and radii; similarly, when $\epsilon$ becomes more negative with $\alpha < 0$, the mass–radius configurations also increase, although the masses are not as large as in the configuration with $\alpha > 0$. 

The parameter $\alpha$ acts on the matter sector. When $\alpha \neq 0$, the effective density and pressure are modified, whereas when $\alpha = 0$, they reduce to the f(R) form. Since $\alpha \neq 0 $, the system becomes more flexible in maintaining hydrostatic equilibrium. As $\epsilon$ becomes more negative, larger $\alpha$ values are required to obtain stable high-mass solutions. As $\alpha$ increases, the additional contribution to the effective pressure also increases. The next case is $\alpha < 0$. Smaller $\alpha$ reduces the $T^2$ term's contribution, resulting in a lower maximum mass than for $\alpha > 0$. In other words, positive values of the coupling parameter $\alpha$ enhance the star’s ability to harbor a large mass, whereas negative values of $\alpha$ weaken the additional effective pressure contribution, leading to smaller stellar masses. 

In the central-pressure picture, GR is the standard case with no contributions from $\epsilon$ or $\alpha$. In modified gravity models, as $\epsilon$ decreases, the stellar structure shows that the central pressure required to reach the maximum mass also decreases, while the corresponding $\alpha$ increases. The numerical results show that increasing $\alpha$ decreases the central pressure required to reach the maximum mass. 

This behavior can be attributed to the combined effects of the modified geometry, which weakens the effective gravitational attraction, and the increased effective pressure, which provides additional pressure support. As a result, the stellar configuration can reach hydrostatic equilibrium at lower central pressures and develop larger masses and radii.

From a compactness perspective, the Buchdahl bound serves as a benchmark for assessing the physical acceptability of stellar configurations. Since the Buchdahl bound states that the upper limit for the compactness of a static, spherically symmetric, perfect fluid object is $\approx 0.444$, all obtained solutions remain below this value. Therefore, although the $f(R,T^2)$ gravity model yields larger stellar masses, the configurations have not yet undergone gravitational collapse. In other words, the quark stars remain stable in these configurations.

\subsection{Parameter Constraints from Observational Compact Objects}
To evaluate the viability of the $f(R,T^2)$ gravity model, this research constrains the parameters $\epsilon$, $\alpha$, and $p_c$ to examine whether the M-R relation of this model can accommodate the observational data of compact objects. This research considers four compact objects, namely HESS J1731-347, which has a relatively low mass of $M = 0.77_{-0.17}^{+0.20} \, M_{\odot}$ and a radius of $R = 10.4_{-0.78}^{+0.86}$ km \cite{Doroshenko:2022nwp}, the secondary component of GW190814 with a mass of $M = 2.6 \, M_{\odot}$ \cite{LIGOScientific:2020zkf,Danarianto:2024vih}, PSR J1614-2230, with a mass of $M = 1.928 \pm 0.017 \, M_{\odot}$ and a radius of $R > 10$ km \cite{Miller2016}, and lastly, PSR J0437-4715, with a mass of $M = 1.418 \pm 0.037 \, M_{\odot}$ and a radius of $R = 11.36_{-0.63}^{+0.95}$ km \cite{Choudhury2024PSRJ0437}. In Figs. \ref{fig:perbandingan} and \ref{fig:mr-fixed}, we compared the corresponding observational data with mass-radius relation results for quark stars using various combinations of $\epsilon$ and $\alpha$.

In the left panel of Fig. \ref{fig:perbandingan}, the model with parameters $\epsilon < 0$ and $\alpha > 0$ satisfies all mass and radius constraints for the observed objects. Specifically, for GW190814.  All possible combinations of the parameters $\epsilon$ and $\alpha$ other than the one in left panel of Fig. \ref{fig:perbandingan} such as the one in right panel of Fig. \ref{fig:perbandingan} as well as the panels in Fig.\ref{fig:mr-fixed} are implemented to examine whether the $f(R,T^2)$ gravity model used can also reach the target mass of $2.6 \, M_{\odot}$ by taking the central pressure at which the maximum mass of the parameter combinations reaches $M = 2.6 \, M_{\odot}$. They cannot reach the $M = 2.6 \, M_{\odot}$ constraint.

From Figs. \ref{fig:perbandingan} and \ref{fig:mr-fixed}, it can also be seen that almost all $f(R,T^2)$ and GR plots pass through the mass-radius range of HESS J1731-347. This may be because HESS J1731-347 lies in the low-mass range, where GR and $f(R,T^2)$ differ little. This indicates that for less extreme $f(R,T^2)$ corrections, the model remains consistent with small-- and medium-mass compact stars. In other words, $f(R,T^2)$ gravity behaves close to GR at low to intermediate masses. Similarly, both GR and $f(R,T^2)$ gravity can reach the other two compact objects, PSR J0437-4715 and PSR J1614-2230, as shown in Figs. \ref{fig:perbandingan} and \ref{fig:mr-fixed}. 
In Fig.~\ref{fig:perbandingan}, some plots barely reach those of PSR J0437$-$4715 because negative values of $ \alpha$ yield smaller masses. Furthermore, GR cannot reproduce the observed secondary mass in the GW190814 event. In conclusion, $f(R,T^2)$ gravity can satisfy the mass constraint for these objects if $\epsilon < 0$ and $\alpha > 0$.

\subsection{Speed of Sound} \label{subsect:cs2}
Here, we discuss in detail the speed of sound of $f(R,T^2)$ theory.
\begin{figure}[h!]
	\centering
	\includegraphics[width=1\linewidth]{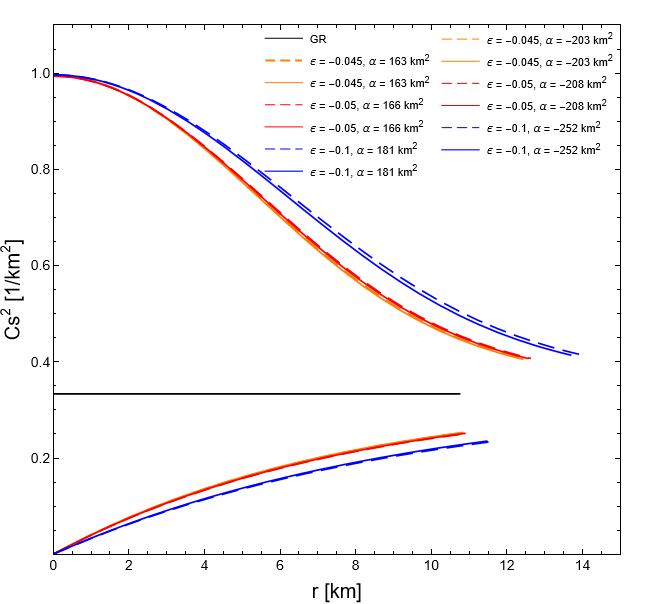}
	\caption{Speed of Sound for Mass-Radius Configuration. Curves approaching 1 correspond to $\alpha > 0$, while curves approaching 0 correspond to $\alpha < 0$.}
	\label{fig:sos}
\end{figure}

From the plots in Fig.~\ref{fig:sos}, all solutions for various values of $\epsilon$ and $\alpha$ satisfy the causality and thermodynamic stability conditions, since $0 \le c_s^{2 \, \mathrm{eff}} \le 1$. As shown in the figure, the GR profile has a nearly constant $c_s^{2 \, \mathrm{eff}}$ value of approximately 0.33. This is consistent with the MIT Bag Model used; since GR does not include any modifications, the derivative yields $dp/d \rho = 1/3$. This result is physical and consistent with the employed equation of state. This value of $1/3$ is the speed of sound of the MIT Bag Model and matches that of QCD's conformal limit. 

For $f(R,T^2)$ gravity, the $c_s^{2 \, \mathrm{eff}}$ value is no longer constant because $\rho$ and $p$ are now expressed in their effective forms. The graph shows two distinct trends for positive and negative values of $\alpha$. Profiles with $\alpha > 0$ tend to have $c_s^{2 \, \mathrm{eff}}$ values that approach 1 and then decrease toward the surface. In contrast, profiles with negative $\alpha$ tend to have $c_s^{2 \, \mathrm{eff}}$ values approaching 0 and then increasing with radius. For $\alpha > 0$, the effective matter becomes very stiff, meaning the effective pressure increases sensitively with changes in effective density, making the stellar interior difficult to compress. This happens because $\alpha > 0$ adds an effective pressure, causing it to rise more rapidly with increasing effective density in the stellar core. The profile then decreases to lower values; this indicates that the stiff region is concentrated at the stellar core, while regions farther from the core become easier to compress and closer to the conformal regime. This occurs because the effective energy density and pressure decrease toward the stellar surface; thus the modified correction becomes smaller. Matter that is difficult to compress implies that the star can sustain a large mass. However, when calculating the effective speed of sound, $p^{\text{eff}}$ and $\rho^{\text{eff}}$ are no longer absolute. Instead, they measure how fast pressure changes relative to how fast density changes. When $\epsilon < 0$, the effective gravitational attraction weakens, and the effective pressure increases, so the pressure response to density variations strengthens, making the effective matter stiffer. 

For the $\alpha < 0$ case, the $c_s^{2 \, \mathrm{eff}}$ value tends to approach 0; this is because a negative $\alpha$ weakens the effective pressure inside the star, meaning that the pressure at the stellar core is not as strong as in the case of a positive $\alpha$. As a result, the matter becomes easier to compress and less stiff. Physically, this means that at the stellar core, the change in pressure relative to the change in energy density is very small, leading to a lower speed of sound. Furthermore, $c_s^{2 \, \mathrm{eff}}$ increases with radius, indicating that the effective matter at the stellar core is more compressible than the outer regions, which become stiffer. This predicts that $\alpha < 0$ weakens the effective pressure response most strongly in the stellar core; meanwhile, as matter is distributed outward, pressure and density decrease, weakening the effective terms so that pressure and density are less affected by the modified correction. Furthermore, when considering the $\epsilon$ parameter, as $\epsilon$ becomes more negative, the $\alpha < 0$ value required to achieve the maximum stellar mass also becomes more negative. More negative $\alpha$ values yield a smaller $p^{\mathrm{eff}}$ and reduce the effective pressure's sensitivity to changes in effective energy density. Consequently, the effective matter at the stellar core becomes more compressible. The plots in Fig.~\ref{fig:sos} also show that smaller $\epsilon$ and larger $\alpha$ make the effective matter stiffer. Conversely, smaller $\epsilon$ and smaller $\alpha$ make the effective matter softer. This indicates that the change in effective pressure relative to the change in effective density depends on the free parameters $\epsilon$ and $\alpha$, as well as on the central pressure of the stellar configuration. 

\subsection{Energy Conditions}
Energy conditions (ECs) assess the physical viability of a star. They ensure that the matter inside the star satisfies the required physical conditions. Subsection \ref{subsect: ecs} presents the ECs used in $f(R,T^2)$ gravity.

\subsubsection{Strong Energy Condition (SEC)}
The Strong Energy Condition (SEC) states that gravity is attractive, causing the focusing of timelike geodesics. 

\begin{figure}[h!]
	\centering
	\includegraphics[width=1\linewidth]{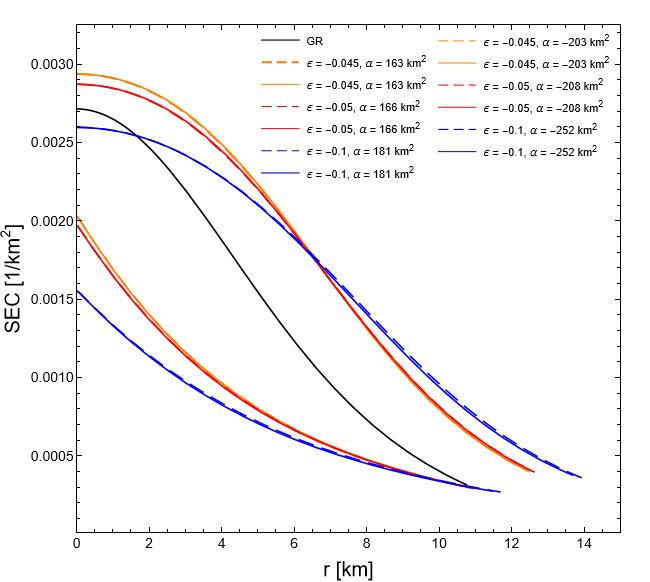}
	\caption{Strong Energy Condition for Mass-Radius Configuration. The upper curves correspond to configurations with $\alpha > 0$, while the lower curves correspond to configurations with $\alpha < 0$.}
	\label{fig:sec}
\end{figure}

Positive SEC values for all M-R profiles indicate that the stellar solutions are physically reasonable. From Fig. \ref{fig:sec},  the combination of $\epsilon = -0.045$ and $\alpha = 163 ~\text{km}^2$ yields the highest SEC value for both the exact and Taylor-expanded forms. This is due to the $+3p^{\text{eff}}$ term in the SEC equation, along with the contribution of $\epsilon$ from geometric corrections in the $f(R)$ sector and its derivatives. The second-highest value corresponds to $\epsilon = -0.05$ and $\alpha = 166~\text{km}^2$. This can be attributed to the fact that these parameter values are relatively close to those of the configuration with $\epsilon = -0.045$ and $\alpha = 163 ~\text{km}^2$. The third-highest value corresponds to GR. This is because Einstein's field equations directly couple curvature to the energy-momentum tensor without any additional modifications. In contrast, in $f(R,T^2)$ gravity, the geometric contributions shift and combine with the matter sector as effective terms, modifying the relation between curvature and matter. From the SEC graph, as $\epsilon$ approaches 0, the SEC value increases in the stellar core. This implies that the effect is governed not only by the large effective pressure contribution from the term $ +3p^ {\mathrm{eff}}$, but also by $\epsilon$. In addition, as $\epsilon$ approaches 0, the required central pressure to reach the maximum mass increases, thereby increasing the $3 p^{\mathrm{eff}}$ factor. 

Under the SEC criterion, higher SEC values indicate greater internal pressure and a stronger tendency toward geodesic focusing. Graphs with larger positive $\alpha$ values show higher SEC values than those with negative $\alpha$ values. This is because the input central pressure required to reach the maximum stellar mass is higher, increasing the contribution from $p^{\mathrm{eff}}$. Conversely, for negative $\alpha$, the input central pressure required to reach the maximum mass is lower, and the contribution from $p^{\mathrm{eff}}$ decreases, resulting in a lower SEC curve. Notice that smaller $\alpha$ values behave differently from the others, showing a sharper decline in their profiles. Physically, as discussed in Subsection \ref{subsect:cs2}, a negative $\alpha$ makes the effective matter softer, so the effective pressure changes more slowly relative to changes in the effective density. This means the density can change significantly without a great corresponding change in pressure. As a result, the matter distribution becomes more compressed at the core and then decreases more rapidly outward than the pressure. 

\subsubsection{Weak Energy Condition (WEC)}

\begin{figure}[h!]
	\centering
	\includegraphics[width=1\linewidth]{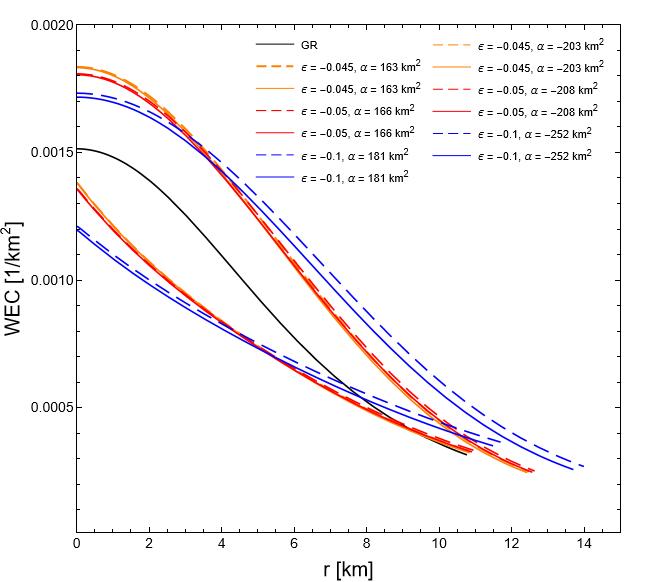}
	\caption{Weak Energy Condition for Mass-Radius Configuration. The upper curves correspond to configurations with $\alpha > 0$, while the lower curves correspond to configurations with $\alpha < 0$.}
	\label{fig:wec}
\end{figure}
The WEC states that an observer measures non-negative energy density. As with the SEC, positive WEC values across all curve profiles indicate that the stellar solutions are physically viable.  From Fig. \ref{fig:wec}, the WEC values are lower than those in the SEC graph. This is influenced by the $\rho^{\mathrm{eff}} + 3p^{\mathrm{eff}}$ factor in the SEC, whereas the WEC only involves $\rho^{\mathrm{eff}}$ and $\rho^{\mathrm{eff}} + p^{\mathrm{eff}}$, meaning that the pressure contribution is not as large as in the SEC. Especially in this case, the factor used to calculate WEC is $\rho^{\mathrm{eff}} \geq 0$ (in this case, the condition $\rho^{\mathrm{eff}} + p^{\mathrm{eff}} \ge 0$ will be represented in the NEC since they share the same requirement). This implies that the WEC is easier to satisfy than the SEC. 

The highest WEC plot is for the values $\epsilon = -0.045$ and $\alpha = 163 ~\text{km}^2$. This implies that this parameter combination yields the maximum energy density at the stellar core. Looking at smaller $\epsilon$ values, the effective gravity effect is weaker, and the effective density distribution decreases more gradually with radius. This could also be because, at higher $\alpha$, the effective matter is stiffer. Although one might directly expect $\rho^{\mathrm{eff}}$ to decrease since $\alpha > 0$, this result indicates that the WEC behavior is determined not only by local corrections to $\rho^{\mathrm{eff}}$ terms but also by global changes in the stellar structure resulting from the solution of the differential equation system. As shown in the figure, higher WEC values at the stellar center are dominated by configurations with $\alpha > 0$, since positive $\alpha$ increases central pressure and allows the star to attain more massive configurations. Higher central pressure also corresponds to higher energy density. Furthermore, the figure shows that the contributions from $p$ and $\rho$ outweigh the negative contribution of $\alpha$ in the correction term for $\rho^{\text{eff}}$. 

In contrast, for $\alpha < 0$, the resulting WEC values are smaller than for $\alpha > 0$ because the effective pressure distribution and energy density are reduced. Locally, the correction term does increase $\rho^{\mathrm{eff}}$. Still, solving the TOV system of differential equations to obtain the M-R curves shows that $\alpha < 0$ tends to reduce the interior effective pressure distribution. As a result, the total $\rho^{\mathrm{eff}}$ and WEC values obtained from the solution of the modified system of differential equations can remain lower than those for $\alpha > 0$. This is why the WEC graphs for $\alpha > 0 $ are higher than for $\alpha < 0$. Similar to the SEC, a large negative $\alpha$ value exhibits different behavior. This is because $\alpha < 0$ weakens the effective pressure and energy density, preventing them from being distributed further. In contrast, a small $\epsilon$ value weakens gravity, thereby causing the WEC to decrease more sharply with radius. Furthermore, in terms of the speed of sound, a more negative $\alpha$ makes the effective matter softer.

As in the SEC case, as shown in Fig.~\ref{fig:wec}, the WEC depends not only on $\alpha$ but also on $\epsilon$ and central pressure. The contribution of $\epsilon$ arises through the geometric corrections introduced by the $f(R)$ and its derivatives.

\subsubsection{Null Energy Condition (NEC)}
\begin{figure}[h!]
	\centering
	\includegraphics[width=1\linewidth]{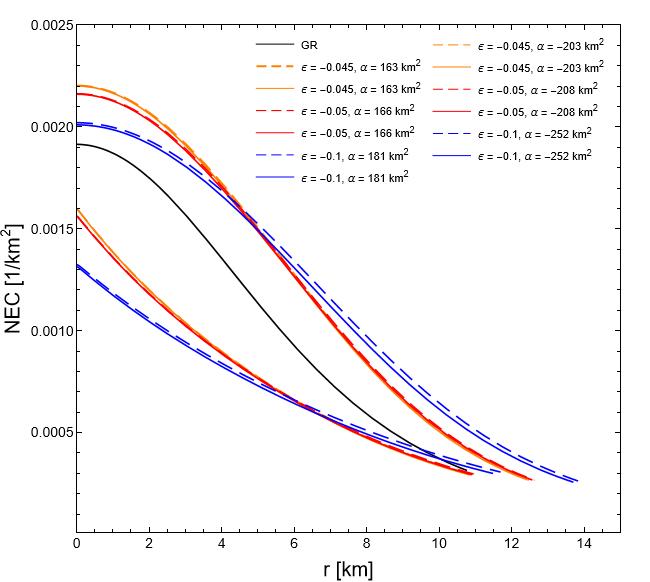}
	\caption{Null Energy Condition for Mass-Radius Configuration. The upper curves correspond to configurations with $\alpha > 0$, while the lower curves correspond to configurations with $\alpha < 0$.}
	\label{fig:nec}
\end{figure}

The NEC requires that the sum of the energy density and pressure along a null trajectory (lightlike direction) must be non-negative. As in the previous cases, a positive NEC value across all curve profiles indicates a physical, stable stellar solution, meaning gravity remains attractive on null geodesics. As shown in Fig. \ref{fig:nec}, the $\rho^{\text{eff}} + p^{\text{eff}}$ term in the NEC cancels the matter correction terms. However, positive $\alpha$ values produce higher central pressures, leading to larger NEC values at the stellar center. Consequently, configurations with positive $\alpha$ dominate the highest NEC values near the stellar core. However, the parameter $\epsilon$ also affects this behavior through the geometric correction terms. For  GR, the contribution comes only from the pressure term $p$ and the energy density $\rho$, with no additional correction terms that enhance its value. As in the SEC explanation, $\alpha > 0$ yields a stiff effective matter. However, based on the speed of sound, the stiffest effective matter has a negative $\epsilon$ and a large positive $\alpha$. As a result, when matter is distributed from the core, the stiffer effective matter can maintain high density out to large radii, raising the graph.

As with the previously discussed ECs, a large negative $\alpha$ behaves differently. For $\alpha \ll 0$, it softens the effective matter. It modifies the stellar structure through the contribution of the $T^2$ term in the field equations, causing $\rho^{\mathrm{eff}}(r)$ and $p^{\mathrm{eff}}(r)$ to decrease more sharply with respect to the radius, since the effective matter is less able to distribute further outward. In contrast, $\alpha > 0$ makes the effective matter stiffer, so the effective pressure responds more sensitively to small changes in the effective energy density, allowing matter to be distributed more widely and resulting in a more gradual decrease. As a result, soft effective matter produces a different trend line than stiff effective matter.

As in the previous case, the NEC depends not only on $\alpha$ but also on $\epsilon$ and central pressure. The parameter $\epsilon$ contributes through the geometric correction terms introduced by the $f(R)$ sector of the gravity model and its derivatives.

\subsubsection{Dominant Energy Condition (DEC)}
\begin{figure}[h!]
	\centering
	\includegraphics[width=1\linewidth]{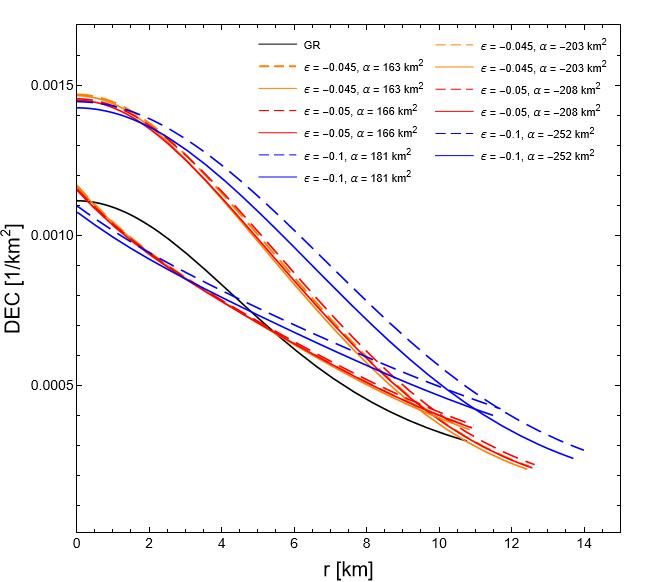}
	\caption{Dominant Energy Condition for Mass-Radius Configuration. The upper curves correspond to configurations with $\alpha > 0$, while the lower curves correspond to configurations with $\alpha < 0$.}
	\label{fig:dec}
\end{figure}

The DEC requires that, for any observer, the energy density must be positive and the energy–momentum flux must be causal, meaning it cannot propagate faster than the speed of light. Under the DEC, the energy density must dominate the pressure. As with the other three ECs, a positive DEC across all profiles indicates a physical, stable stellar solution. In GR, neither geometric nor matter modifications contribute; therefore, pressure and energy density arise solely from pure quark matter rather than being affected by modified gravity. To sustain the stars against gravity, the pressure must be sufficiently large, since both gravity and the energy density are also large. As seen in Fig. \ref{fig:dec}, when $\epsilon$ approaches 0, gravity becomes more attractive, thereby requiring greater pressure to sustain the stars against gravitational collapse. However, the DEC depends not only on $\alpha$ but also on central pressure and $\epsilon$. $\epsilon$ enters through the geometric correction terms associated with the $f(R)$ form and its derivatives. As mentioned in the WEC discussion, $\rho$ and $p$ are not fixed constants but functions of $r$ obtained from solutions to the modified system of differential equations. Based on the DEC profiles, the effective energy density dominates the effective pressure.

The same happens when $\alpha < 0$. As $\epsilon$ approaches 0 while $\alpha < 0 $, the gravitational attraction becomes stronger. As a result, greater pressure is needed to support a massive star against collapse; however, when $\alpha < 0$, the central pressure and energy density required to achieve a large stellar mass are also lower, which tends to reduce the DEC values, particularly near the stellar center. As $\alpha$ becomes more negative, $p^{\mathrm{eff}}$ decreases, making $\rho^{\mathrm{eff}}$ more dominant, especially near the stellar surface. As a result, the trend line's shape changes little compared to the WEC. The condition $\alpha < 0$ also shows distinct behavior, yielding lower effective pressure than configurations with $\alpha > 0$. When the effective gravity is stronger, the pressure support is less able to maintain the matter distribution out to large radii. Thus, when the matter is distributed outside the core, the DEC profile decreases more rapidly. As a result, negative $\alpha$ values soften the effective matter and reduce the effective pressure, causing the DEC profile to decrease more rapidly with increasing radius.

\section{Conclusion}

This work investigates the mass-radius structure and stability of isotropic quark stars within the $f(R,T^2) = R^{1+\epsilon} + \alpha \kappa^2 T_{\mu \nu} T^{\mu \nu}$ model using the MIT Bag Model EoS, by constraining the parameters $\epsilon$ and $\alpha$. $f(R,T^2)$ gravity can predict larger masses, radii, and compactness than GR. The parameters $\epsilon$ and $\alpha$ strongly affect predictions of the star's maximum mass, radius, and compactness. The results indicate that a more negative $\epsilon$ increases geometric nonlinearity and increases deviation from GR, thereby reducing the tendency toward gravitational collapse. Furthermore, the results for $\alpha$ indicate that larger $\alpha$ increases the effective pressure within stars; thus, the combination of $\epsilon$ and $\alpha$ enables stars to attain larger mass and radius. The comparison between the configurations obtained from the exact model and the Taylor expansion shows only minor differences when $\epsilon$ approaches zero. However, as $\epsilon$ deviates further from zero, the difference between the two configurations becomes increasingly significant. The constrained combination of $\epsilon$ and $\alpha$ can also accommodate observational data from compact objects such as HESS J1731-347, PSR J0437-4715, and PSR J1614-2230, and especially GW190814, which GR cannot explain. The parameter combinations $\epsilon = -0.1$ and $\alpha = 181~ \text{km}^2$, and $\epsilon = -0.05$ and $\alpha = 166 ~ \text{km}^2$, both using the exact model and the Taylor expansion, can simultaneously satisfy the observational constraints from these compact objects.

The speed-of-sound analysis shows that all configurations satisfy the causality condition and thermodynamic stability, ensuring the solutions are stable. In GR, the speed of sound equals $0.33$, consistent with the MIT Bag Model and the conformal limit for quark matter. Generally, configurations with $\epsilon < 0$ and $\alpha > 0 $ lead to higher $c_s^{2 \, \mathrm{eff}}$, meaning that the effective matter becomes stiffer, whereas configurations with $\epsilon < 0$ and $\alpha < 0$ generally produce lower $c_s^{2 \, \mathrm{eff}}$, meaning that the effective matter becomes softer and more compressible.

Every resulting mass-radius configuration satisfies the SEC, WEC, NEC, and DEC conditions, indicating that the solutions are physically viable. Overall, the energy conditions cannot be decided from $\rho^{\text{eff}}$ and $p^{\text{eff}}$ because there is also a contribution from the parameter $\epsilon$ through the geometric correction arising from the $f(R)$ sector of the model, which also plays a role in determining the final result.


\acknowledgments
	We thank Fahmi Fauzi and Muhammad Hafizt Al Ghifari for the useful discussion. In this work, we are supported by Hibah Fundamental Directorate General of Higher Education (DIKTI), PKS-178/UN2.RST/HKP.05.00/2026.

\section*{Statement of the Conflict of Interest}

All authors hereby declare that we do not have any conflict of interest.

\section*{Data Availability Statement}

Data sharing is not applicable to this article as no data sets were generated or analyzed during the current study.





\nocite{*}


\end{document}